\documentclass[journal]{IEEEtran} 
\usepackage{cite}
\usepackage[linesnumbered,ruled,vlined]{algorithm2e}
\usepackage{algpseudocode}
\usepackage[small,bf]{caption}
\usepackage[cmex10]{amsmath}
\usepackage{amssymb,latexsym,epsfig,subfigure,epic,amscd,mathrsfs,euscript,eufrak,bm}
\usepackage{color}
\usepackage{booktabs}% http://ctan.org/pkg/booktabs
\usepackage{array}% http://ctan.org/pkg/array
\usepackage{bbm}

\usepackage{amsfonts}
\usepackage{amsopn}
\usepackage{graphicx}
\usepackage{url} 
\usepackage{empheq}
\usepackage[inline]{enumitem}

\newtheorem{theorem}{\bf{Theorem}}
\newtheorem{remark}{\bf{Remark}}

\newcommand{\subscript}[2]{$#1 _ #2$}

\DeclareMathAlphabet\mathbfcal{OMS}{cmsy}{b}{n}

\newcommand{\bs}[1]{ \ensuremath{ \boldsymbol{#1} }}

\makeatletter
\def\blfootnote{\xdef\@thefnmark{}\@footnotetext}
\makeatother

\title{Distributed Sequential Hypothesis Testing with Dependent Sensor Observations}
\author{Shan Zhang$^{\ast}$, Prashant Khanduri$^{\ast}$, and Pramod K. Varshney$^{\ast}$\\
	$^{*}$Department of Electrical Engineering and Computer Science, Syracuse University, Syracuse, New York-13244\\ 
szhang60@syr.edu,	pkhandur@syr.edu, varshney@syr.edu  
\thanks{The work was supported by Air Force
Office of Scientific Research under Grants FA9550-16-1-0077 and FA9550-17-1-0313 under the DDDAS
program. }
}
\begin{document}

\maketitle

\begin{abstract}
In this paper, we consider the problem of distributed sequential detection using wireless sensor networks (WSNs) in the presence of imperfect communication channels between the sensors and the fusion center (FC). We assume that sensor observations are spatially dependent. We propose a copula-based distributed sequential detection scheme that characterizes the spatial dependence. Specifically, each local sensor collects observations regarding the phenomenon of interest and forwards the information obtained to the FC over noisy channels. The FC fuses the received messages using a copula-based sequential test. Moreover, we show the asymptotic optimality of the proposed copula-based sequential test. Numerical experiments are conducted to demonstrate the effectiveness of our approach.
\end{abstract}
\noindent

\section{Introduction}
\label{sec:i}
Wireless sensor networks (WSNs) consist of a large number of spatially distributed sensors collaborating to solve inference problems, such as detection, estimation, and classification \cite{cheng2012survey}. The local sensors collect noisy observations of the phenomenon of interest, and transmit the collected information to the fusion center (FC) which fuses the received data and produces a global decision. In this paper, we study the problem of distributed detection, where local sensors and the FC collaborate to detect the presence or absence of the target of interest.
  
The distributed detection problems in sensor networks with \textit{fixed-sample-size} (FSS) have been studied extensively \cite{veeravalli2012distributed, Varshney_Book_1997}, where the goal often is to minimize some cost at the FC based on a fixed number of observations. However, many detection problems are inherently sequential, where the observations are collected and processed sequentially \cite{tartakovsky2014sequential}. In addition, sequential decision procedures provide additional advantage of faster decision making. In \cite{wald1945sequential}, Wald proposed the sequential probability ratio test (SPRT) for sequential binary hypothesis testing. SPRT was shown to be optimal in the sense that it takes minimum time on average to make a decision among all the tests which guarantee the same probabilities of error. Distributed versions of SPRT were first proposed in \cite{veeravalli1993decentralized,Veeravalli_JFI_1999,Mei_TIT_2008}, where the proposed algorithms were shown to be (asymptotically) optimal under independence assumptions of the sensor observations. Moreover, in \cite{veeravalli1993decentralized,Veeravalli_JFI_1999,Mei_TIT_2008}, the channels from the sensors to the FC were assumed to be noiseless. In this work, we propose a novel approach for sequential detection in a distributed sensor network which does not require the observations to be independent across sensors. Also, the channels from the sensors to the FC can be noisy.  

Dependence across spatially distributed sensors is a critical issue in distributed detection problems. This dependence exists due to a variety of reasons such as sensing of the same phenomenon and corruption by correlated noise. Some works have studied the effect of dependence on the performance of detection systems \cite{drakopoulos1991optimum, kam1992optimal, willett2000good, kasasbeh2016hard, kasasbeh2017soft, Khanduri_ICASSP_2017}. However, in many systems of interest, this dependence is ignored and the sensor observations  are generally assumed to be independent. Ignoring this underlying dependence degrades detection performance as shown in \cite{iyengar2011parametric, he2012fusing, zhang2019fusion}. One way of combating this unknown dependence is by posing the sequential detection problem in a non-parametric framework \cite{Khanduri_ICASSP_2018,Khanduri_Allerton_2018}. However, providing optimality guarantees for such problems is often not feasible. In this work, we use a copula-based approach to model the dependence and show its asymptotic optimality. 

Copula-based approach \cite{nelsen2013introduction} is a flexible parametric methodology that decomposes the joint distribution of sensor observations into arbitrary marginal distributions and a multivariate distribution (dependence) that is referred to as the 
\textit{copula} distribution. It has been shown that copula-based fusion of multiple sensing observations can significantly improve inference performance \cite{iyengar2011parametric, he2012fusing, sundaresan2011copula, zhang2019fusion}. %In \cite{sundaresan2011copula, zhang2019fusion}, the authors proposed copula based fusion rules for distributed detection in a FSS setup. 
%Moreover, in \cite{zhang2019fusion}, for high-dimensional dependence structures, a regular vine (R-Vine) copula based fusion rule was proposed at the FC, where the multivariate copula was constructed by a set of bivariate copulas.
However, the work in \cite{iyengar2011parametric, he2012fusing, sundaresan2011copula, zhang2019fusion} were FSS based tests.
In contrast, in this paper, we propose a novel copula-based distributed sequential hypothesis test with the main contributions of the work summarized below. %We show that the proposed test is not only asymptotically optimal, but also captures the unknown %(potentially high dimensional) 
%dependence across the sensors. Via simulations, the proposed test is shown to outperform the SPRTs which usually ignore the underlying unknown dependence across sensors. Moreover, we have shown that the copula-based SPRT is less sensitive to low signal-to-noise ratios (SNRs).
%We summarize our contributions as follows. 
\begin{itemize}[leftmargin = 3.5 mm]
	\item We propose a copula-based sequential hypothesis testing approach to model the dependence across sensors with imperfect communication from the sensors to the FC.
	\item We show that the proposed test is not only asymptotically optimal, but also captures the unknown %(potentially high dimensional) 
dependence across the sensors.

	\item Via simulations, we show that the proposed copula-based SPRT outperforms the SPRTs which usually ignore the underlying unknown dependence across sensors. Also, we show that the copula-based SPRT is less sensitive to low signal-to-noise ratios (SNRs).
\end{itemize}

The rest of the paper is organized as follows. In Section \ref{sec:CT}, we provide a brief introduction to copula theory. %including standard multivariate copulas and regular vine copulas. 
In Section \ref{sec:PF}, we introduce the distributed sequential detection system. In Section \ref{sec:cen}, we propose a copula-based SPRT. In Section \ref{sec:nr}, we demonstrate the effectiveness of the proposed distributed sequential scheme through numerical examples. Finally, in Section \ref{sec:conclusion}, we conclude the paper.

\section{Copula Theory Background}
\label{sec:CT}
A copula is a multivariate distribution with uniform marginal distributions, and it characterizes the dependence among multiple continuous variables. 
The unique correspondence between a multivariate copula and any multivariate distribution is stated in Sklar's theorem \cite{nelsen2013introduction} which is a fundamental theorem that forms the basis of copula theory.    

\begin{theorem}[Sklar's Theorem]
\label{theorem:s}
For random variables $x_1,\ldots,x_d$, their joint distribution function $F$ can be cast as
\begin{equation}
\label{CopEq1}
F(x_1,x_2,\ldots,x_d) = C(F_1(x_1),F_2(x_2),\ldots,F_d(x_d)),
\end{equation}
where $F_1,F_2,\ldots,F_d$ are continuous marginal Cumulative Distribution Functions (CDFs) for all the random variables and $C$ is a unique $d$-dimensional copula. Conversely, given a copula $C$ and univariate CDFs $F_1,\ldots,F_d$, $F$ in \eqref{CopEq1} is a valid multivariate CDF with marginals $F_1,\ldots,F_d$.
\end{theorem}

For absolutely continuous distributions $F$ and $F_1,\ldots,F_d$, the joint Probability Density Function (PDF) of random variables $x_1,\ldots,x_d$ can be obtained by differentiating both sides of \eqref{CopEq1}: \begin{equation}
\label{CopEq2}
f(x_1,\ldots,x_d)\! = \! \Big(\prod_{m=1}^{d}f_m(x_m)\Big)c(F_1(x_1),\ldots,F_d(x_d) | \bs \phi),
\end{equation}
where $f_1, \ldots, f_d$ are the marginal densities and $c$ is referred to as the density of the multivariate copula $C$, that is given by 
\begin{equation}
\label{CopDens}
c(\mathbf F(\mathbf x) | \bs \phi) = \frac{\partial^d C(\mathbf F(\mathbf x) | \bs \phi)}{\partial F_1,\ldots,\partial F_d},
\end{equation}
where $\mathbf F(\mathbf x) = [F_1(x_1), \ldots, F_d(x_d)]$. Moreover, $\bs{\phi}$ is the \textit{dependence parameter} that characterizes the amount of dependence among $d$ random variables. Typically, $\bs{\phi}$ is unknown a \textit{priori} and needs to be estimated, e.g., using Maximum Likelihood Estimation (MLE) or Kendall's $\tau$~\cite{he2015heterogeneous}. Also, in general, $\bs{\phi}$ may be a scalar, a vector or a matrix.

%Thus, given specific univariate marginal distributions $F_1,\ldots,F_d$ and copula model $C$, the joint distribution function $F$ can be constructed by
%\begin{equation}
%F\left(F_1^{-1}(F_1(x_1)), \ldots, F_d^{-1}(F_d(x_d))\right) = C(\mathbf F(\mathbf x)),
%\end{equation}
%where $F_m^{-1}(\cdot)$ are the inverse distribution functions of the marginals, $m = 1, 2, \ldots, d$.

Note that $C(\cdot)$ is a valid CDF and $c(\cdot)$ is a valid PDF for uniformly distributed random variables $F_m$, $m = 1, 2, \ldots, d$. Since $F_m$ represents the CDF of $x_m$, the CDF of $F_m$ naturally follows a uniform distribution over $[ 0,1]$. 

Various families of multivariate copula functions are described in \cite{nelsen2013introduction}, such as Elliptical and Archimedean copulas. Since different copula functions may model different types of dependence, selection of the optimal copula model to fit given data is a key problem. Moreover, note that the copula based measure of dependence is independent of marginals \cite{calsaverini2009information}.

\section{Problem Formulation}
\label{sec:PF}
\begin{figure}
	\centering
	\includegraphics[height=2 in,width=3 in]{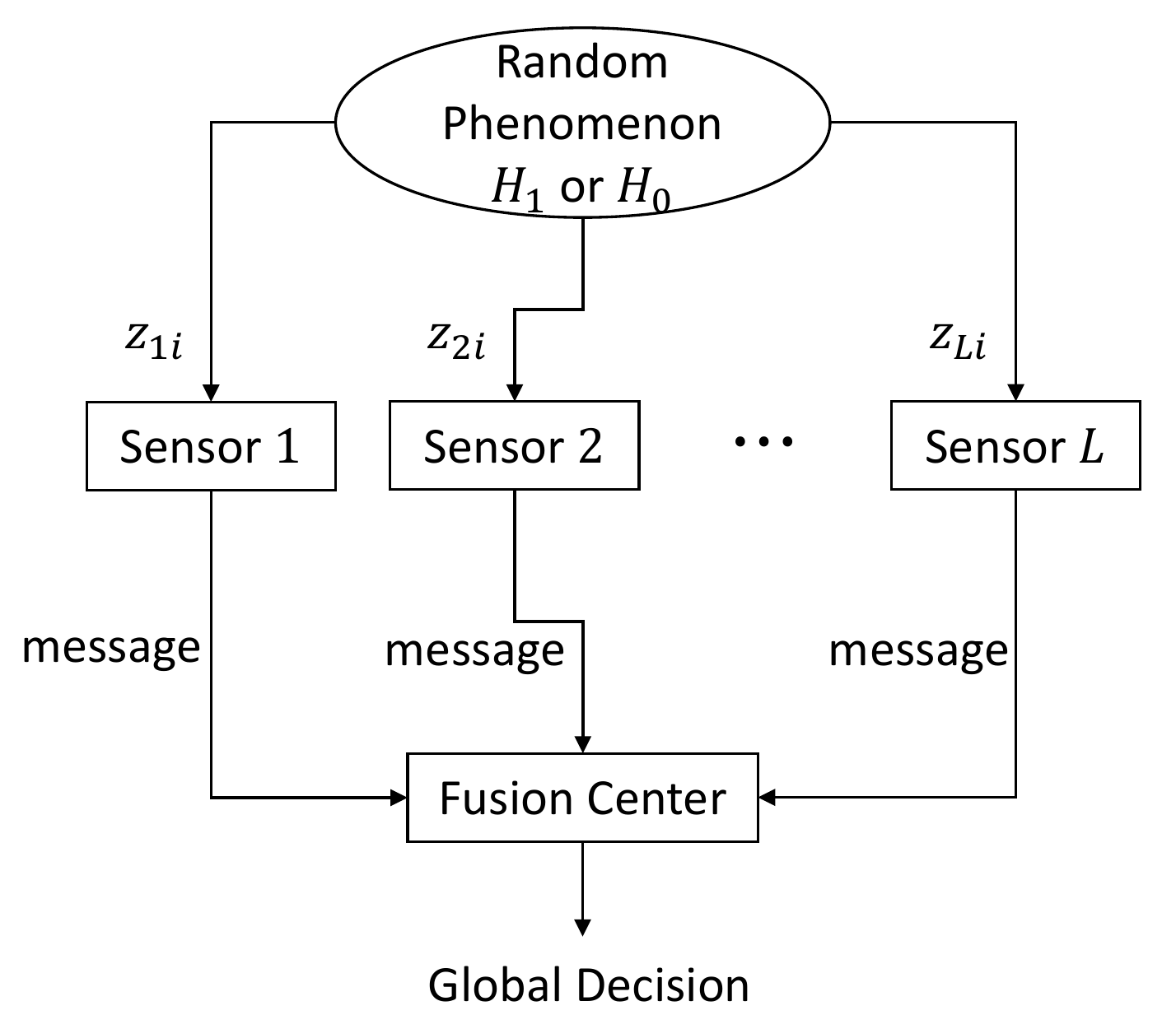}
	\caption{Parallel distributed detection system.}
	\label{fig:system}
\end{figure}

Consider a sequential binary hypothesis testing problem for the sensor network shown in Fig. \ref{fig:system}. The two hypotheses, denoted by $H_1$ and $H_0$, are associated with the random phenomenon of interest that is monitored by $L$ sensors. Suppose that the $l$th sensor acquires observations $z_{li}, l = 1, 2, \ldots, L$ at each time instant $i = \{1,2,\ldots\}$, and forwards the observations over noisy channels to the FC that runs a sequential test and produces a global decision based upon its received messages from the sensors. %The channels from the sensors to the FC are corrupted by additive noise $w_{li}, l = 1, 2, \ldots, L, i = 1, 2, \ldots$.
%Instead of transmitting the raw observations, each local sensor conducts a sequential test and sends binary decisions $U_{li}, l = 1, 2, \ldots, L, i = 1, 2, \ldots$ to the FC sequentially. 
At the FC, the sequential procedure has three possible outcomes: it may either 1) accept $H_0$ and stop the testing, or 2) accept $H_1$ and stop the testing or 3) make no decision and acquire a new observation. The FC repeats this process until a decision is reached, in which case the test stops. Let $T$ denote the stopping time. The goal is to minimize the expected stopping time $\mathbb E_k [T]$ under hypothesis $k = 0,1$ given that $P_F \leq \alpha, P_M \leq \beta$, where $P_F$ is the probability of false alarm with constraint $\alpha \in (0, 1/2)$ and $P_M$ is the probability of miss detection with constraint $\beta \in (0, 1/2)$. We first make the following assumptions. 

\begin{itemize}[leftmargin = 3.5 mm]
\item Sensor observations $z_{li}, l = 1, 2, \ldots, L, i = 1, 2, \ldots$ are continuous random variables and independent and identically distributed (i.i.d.) over time. 
%\item The PDFs of sensor observations $z_{li}, l = 1, 2, \ldots, L$, namely the marginal PDFs, are known under both hypotheses and given by $g_{k, l}(z_{li}), k = 0, 1$. 
\item The channel links between the sensors and the FC are corrupted by additive noise $w_{li}, l = 1, 2, \ldots, L, i = 1, 2, \ldots$. Also, $w_{li}, l = 1, 2, \ldots, L$ are assumed to be i.i.d. over time and independent of the messages sent by the local sensors. 
\item The signal received at the FC corresponding to sensor $l$ at time $i$, after being corrupted by the imperfect channel, is $y_{li} = z_{li} + w_{li}$.
\item The marginal PDFs $f_{k, l}(y_{li})$ and CDFs $F_{l}^k(y_{li})$ with $k \in \{0, 1\}$ under both hypotheses are assumed to be known. However, the manifestation of dependence among $y_{1i}, \ldots, y_{Li}$, namely the copula density function $c_k(\cdot | \bs{\phi}_{k}), k = 0, 1$ (see \eqref{CopEq2}), is not available \textit{a priori}. This dependence may result from the dependent messages that local sensors sent, the dependent additive noise or both.
\end{itemize}

%\begin{remark}
%In some applications, we may not have any prior information related to the phenomenon of interest, namely, the marginal PDFs and marginal CDFs are not known. However, the marginal PDFs can be estimated non-parametrically using kernel density estimators \cite{wassermann2006all} that provide a smoothed estimate of true density by choosing the optimal bandwidth so that an accurate estimate is obtained. \textcolor{blue}{Moreover, the marginal CDF $F_l^k(\cdot)$ for sensor $l, l = 1, \ldots, L$ under hypothesis $k, k = 0, 1$ can be obtained by using Empirical Probability Integral Transforms (EPIT) \cite{he2012fusing}:
%\begin{equation}
%\hat{F}_l^k(\cdot) = \frac{1}{N} \sum_{j=1}^{N} \mathbf{1}_{y_{lj} < \{ \cdot \}},
%\label{empCDF}
%\end{equation}
%where $\mathbf{1}_{\{\cdot\}}$ is the indicator function and $N$ is the number of observations for the estimation.}
%\end{remark}

Before we proceed, we recall the definition of Kullback-Leibler (KL) divergence between two PDFs $f(x)$ and $g(x)$, denoted by $D(f(x) || g(x))$, and given as 
$$D(f(x) || g(x)) = \int f(x) \text{log} \left ({f(x)}/{g(x)} \right) dx.$$ 

Moreover, for any PDFs $f(x)$ and $g(x)$, $D \left(f(x) || g(x) \right) \geq 0$ with equality if and only if $f(x) = g(x)$. Throughout this paper, we assume that
%\begin{enumerate}[label=(\subscript{A}{{\arabic*}})]
\begin{enumerate}[label=\subscript{A}{\arabic*}:]
\item For each sensor $l$, $D \left (f_{0, l}(\cdot) || f_{1, l}(\cdot) \right)$, $D \left(c_0(\cdot | \bs{\phi}_{0}) || c_1(\cdot | \bs{\phi}_{1})\right)$, $D \left(f_{1, l}(\cdot) || f_{0, l}(\cdot) \right)$ and $D \left(c_1(\cdot | \bs{\phi}_{1}) || c_0(\cdot | \bs{\phi}_{0})\right)$  are finite and positive. 
\item Also, \vspace{-0.2 in}
%\begin{align*}
%%&0 < \sum_{l = 1}^{L}D(f_{0, l}(\cdot | H_0) || f_{1, l}(\cdot | H_1)) < \infty, \\
%&0 < D \left(c_0(\cdot | \bs{\phi}_{0}) || c_1(\cdot | \bs{\phi}_{1})\right)  < \infty, \\
%%&0 < \sum_{l = 1}^{L}D(f_{1, l}(\cdot | H_1) || f_{0, l}(\cdot | H_0)) < \infty, \\
%&0 < D \left(c_1(\cdot | \bs{\phi}_{1}) || c_0(\cdot | \bs{\phi}_{0})\right)  < \infty.
%\end{align*}

\begin{small}
 \begin{align*}
& \int \left (\text{log} \frac{f_{0, l}(\cdot)}{f_{1, l}(\cdot)} \right)^2 f_{0, l}(\cdot) d{(\cdot)}   < \infty, \\
& \int \left (\text{log} \frac{f_{1, l}(\cdot)}{f_{0, l}(\cdot)} \right)^2 f_{1, l}(\cdot) d{(\cdot)} < \infty, \\
& \int \left (\text{log} \frac{c_0(\cdot | \bs{\phi}_{0})}{c_1(\cdot | \bs{\phi}_{1})}\right )^2 c_0(\cdot | \bs{\phi}_{0})d{(\cdot)}   < \infty, \\
& \int \left (\text{log} \frac{c_1(\cdot | \bs{\phi}_{1})}{c_0(\cdot | \bs{\phi}_{0})}\right )^2 c_1(\cdot | \bs{\phi}_{1})d{(\cdot)} < \infty.
\end{align*}
\end{small}
\end{enumerate}
\begin{remark}
Note that the two conditions $A_1$ and $A_2$ guarantee that the two hypotheses are distinguishable, i.e.,  $f_1(\cdot)$ is not equal to $f_0(\cdot)$ almost everywhere.
\end{remark}

\section{Copula-based SPRT}
\label{sec:cen}
In this section, we propose a copula-based sequential test. The corrupted messages $y_{li}, l = 1, 2, \ldots, L, i = 1, 2, \ldots$ are received at the FC sequentially. The FC performs a copula-based SPRT to make a final decision. Specifically, we solve the following binary hypothesis testing problem at the FC: 
\begin{equation}
\label{eq:h1h0}
\begin{split}
H_0  &:  \mathbf y_i  \sim f_0(\mathbf y_i), i = 1, 2, \ldots, \\ 
H_1  &:  \mathbf y_i  \sim f_1(\mathbf y_i), i = 1, 2, \ldots, 
\end{split}
\end{equation}
where $\mathbf y_i = [y_{1i}, \ldots, y_{Li}]$, $f_1$ and $f_0$ denote the joint PDFs under $H_1$ and $H_0$, respectively. Note that due to the existing complex dependence among the received observations, $f_k(\mathbf y_i) \neq \prod_{l=1}^{L}f_{k, l}(y_{li}), k = 0, 1$. %This dependence can be non-linear or even more complex. Moreover, the dependence structure tends to varying over time. 
We take the dependence into account and design a copula-based SPRT at the FC. 

Using Sklar's theorem, i.e., Theorem \ref{theorem:s}, the joint PDFs in \eqref{eq:h1h0} for $i = 1, 2, \ldots, n$ can be expressed in terms of the marginal distributions and the copula densities, $c_1$ and $c_0$, respectively, under $H_1$ and $H_0$ as 
\begin{align}
\label{eq:h1h02}
f_0(\mathbf y) = &\prod_{i=1}^{n} \prod_{l=1}^{L}f_{0, l}(y_{li}) \times c_0(\mathbf F^{0}(\mathbf y_i) | \bs{\phi}_{0}), \\ \nonumber
f_1(\mathbf y) = &\prod_{i=1}^{n} \prod_{l=1}^{L}f_{1, l}(y_{li}) \times c_1(\mathbf F^{1}(\mathbf y_i) | \bs{\phi}_{1}), \nonumber
\end{align}
where $f_{k, l}(y_{li}), l = 1, \ldots, L$ are the marginal PDFs and $\mathbf F^{k}(\mathbf y_i) = [F_1^{k}(y_{1i}), \ldots, F_L^{k}(y_{Li})]$ are the marginal CDFs for all the sensors at time instant $i$ under hypothesis $k, k = 0, 1$. Moreover, $\bs{\phi}_{0}$ and $\bs{\phi}_{1}$ are the parameters of copula $c_0$ and $c_1$ at time instant $i$, respectively. For known $c_0(\cdot | \bs{\phi}_{0})$ and $c_1(\cdot | \bs{\phi}_{1})$, the copula-based SPRT follows the following procedure: for $n = 1, 2, \ldots$,
\begin{equation}
\label{eq:sprt}
\begin{cases}
\Lambda^{n}(\mathbf{y}) \geq A, & \text{decide} \,\, H_1,\\
\Lambda^{n}(\mathbf{y}) \leq -B, & \text{decide} \,\, H_0,\\
-B < \Lambda^{n}(\mathbf{y}) < A, & \text{take another observation},
\end{cases}
\end{equation}
where $A$ and $-B$ are the upper and lower thresholds, respectively, which are predetermined constants such that $P_F \leq \alpha$ and $P_M \leq \beta$. Also, $\Lambda^{n}(\mathbf{y})$ is given as 
\begin{align} \label{eq:lambda2}
\Lambda^{n}(\mathbf{y})  = &\sum_{i=1}^{n} { \sum_{l=1}^L{ \log \frac{f_{1, l}(y_{li})}{f_{0,l}(y_{li})} }} + \sum_{i=1}^{n}{ \log \frac{c_1(\mathbf F^{1}(\mathbf y_i) | \bs{\phi}_1)}{c_0(\mathbf F^{0}(\mathbf y_i) | \bs{\phi}_0)} }.
\end{align}
In general, for given $\alpha$ and $\beta$, exact analytical expressions of the optimal thresholds $A$ and $-B$ are intractable. One may use the approximated thresholds obtained from Wald's SPRT \cite{wald1945sequential}, which are given by 
\begin{equation}
\label{eq:thre}
A \approx \log \frac{1-\beta}{\alpha}, \quad - B \approx \log \frac{\beta}{1-\alpha},
\end{equation}
where if $\alpha, \beta \in (0, {1}/{2})$, we have $-B < A$.

Let $T$ be the stopping time for the proposed sequential scheme in \eqref{eq:sprt}.  Since the messages received at the FC are i.i.d., we have $P(T < \infty | H_k) = 1, k = 0, 1$ under conditions $A_1$ and $A_2$ \cite[Lemma\,3.1.1]{tartakovsky2014sequential}.
%The following Proposition shows that the test stops at finite time with probability one. 
%
%\begin{myprop}
%\label{lemma:1}
%For the centralized copula-based SPRT in \eqref{eq:sprt}, we have
%\begin{equation}
%P(N < \infty | H_k) = 1 \,\, \text{for} \,\, k = 0,1,
%\end{equation}
%under the conditions $A_1$ and $A_2$.
%\end{myprop}
%\textbf{Proof}: See Appendix\ \ref{appendix:lemma1}. \hfill $\blacksquare$
% 
% \begin{remark}
% If $\alpha$ and $\beta$ are sufficiently small, we have $A > 0$ and $B > 0$. 
% \end{remark}
 The goal for the above proposed copula-based SPRT is to minimize the average stopping time $\mathbb E_k[T], k = 0,1$ such that $P_F \leq \alpha, P_M \leq \beta$. 
In Theorem \ref{thm1}, we show its asymptotic optimality as $A, B \to \infty$.
\begin{theorem}
\label{thm1}
For the proposed copula-based SPRT in \eqref{eq:sprt}, as $A \to \infty$ and $B \to \infty$, we have $P_F \approx e^{-A}$ and $P_M \approx e^{-B}$. Moreover, the average stopping time under the two hypotheses is given by 
 \begin{equation}
 \label{eq:stop11}
 \mathbb E_{H_0} [T] \approx \frac{B}{D_0}, \quad
 \mathbb E_{H_1} [T] \approx \frac{A}{D_1},
 \end{equation}
as $A, B \to \infty$, and where
 \begin{align*}
 D_0 &= \sum_{l = 1}^{L}D\left (f_{0, l}(\cdot) || f_{1, l}(\cdot)\right ) + D\left (c_0(\cdot | \bs{\phi}_{0}) || c_1(\cdot | \bs{\phi}_{1})\right ),
 \end{align*}
 and
 \begin{align*}
 D_1 &= \sum_{l = 1}^{L}D \left (f_{1, l}(\cdot) || f_{0, l}(\cdot) \right) + D \left(c_1(\cdot | \bs{\phi}_{1}) || c_0(\cdot | \bs{\phi}_{0}) \right). 
 \end{align*}
 \end{theorem}
\textbf{Proof}: The proof is omitted because of space limitations. $\blacksquare$

{\renewcommand{\arraystretch}{1.2}
\begin{table*}[t]
\centering
\begin{tabular}{|c|c|c|c|c|c|c|c|c|}
\hline
& \multicolumn{4}{c|}{$SNR = -6$ dB} & \multicolumn{4}{c|}{$SNR = -9$ dB}  \\ \cline{2-9} \hline
&\multicolumn{2}{c|}{Product-based SPRT} &\multicolumn{2}{c|}{Copula-based SPRT}&\multicolumn{2}{c|}{Product-based SPRT} &\multicolumn{2}{c|}{Copula-based SPRT} \\
\hline
$\beta$ & $P_F$   & $P_M$ & $P_F$ & $P_M$ & $P_F$   & $P_M$ & $P_F$ & $P_M$   \\ \hline
0.3            & 0.0078 & 0.4341  & 0.0052 & 0.0321 &0.0089 &0.5106 &0.0052 &0.0299 \\ \hline
0.2              & 0.0075 & 0.3786  &0.0039 &0.0186 &0.0080  &0.4366 &0.0046 &0.0205 \\ \hline
0.1             & 0.0056 & 0.2849  &0.0036 & 0.0101  &0.0079  &0.3362 &0.0039 &0.0099   \\ \hline
0.01            & 0.0061 & 0.1245  &0.0030 &0.0010 &0.0077  &0.1416 &0.0037 &8.8000e-4 \\ \hline
0.001            & 0.0060 & 0.0548  &0.0032 &1.1500e-04  &0.0076 &0.0617&0.0037 &8.4000e-5  \\ \hline
\end{tabular}
\caption{Known-copula: Estimated $P_F$ and $P_M$ with $\alpha = 0.01$.} \label{table:eq1}
\end{table*}}

{\renewcommand{\arraystretch}{1.2}
\begin{table*}[t]
\centering
\begin{tabular}{|c|c|c|c|c|c|c|c|c|}
\hline
& \multicolumn{4}{c|}{$SNR = -6$ dB} & \multicolumn{4}{c|}{$SNR = -9$ dB}  \\ \cline{2-9} \hline
&\multicolumn{2}{c|}{Product based SPRT} &\multicolumn{2}{c|}{Copula based SPRT} &\multicolumn{2}{c|}{Product-based SPRT} &\multicolumn{2}{c|}{Copula-based SPRT} \\
\hline
$\alpha$ & $P_F$   & $P_M$ & $P_F$ & $P_M$ & $P_F$   & $P_M$ & $P_F$ & $P_M$  \\ \hline
0.3            & 0.1803 & 0.0918  & 0.1093 & 0.0018  &0.2337 &0.0914 &0.1207 &0.0010 \\ \hline
0.2              & 0.1178 & 0.0968  &0.0702 &6.000e-4 &0.1568 &0.1027  &0.0825  &0.0011 \\ \hline
0.1             & 0.0573 & 0.1019  &0.0366 & 0.0010  &0.0771 &0.1151 &0.0368    &0.0012  \\ \hline
0.01            & 0.0061 & 0.1245  &0.0030 &0.0010  &0.0077 &0.1416 &0.0037 &8.8000e-4 \\ \hline
0.001            & 5.5100e-04 & 0.1352  &3.1100e-4 &9.7100e-04 &8.2200e-4 &0.1563  &3.6900e-4 &8.3000e-3   \\ \hline
\end{tabular}
\caption{Known-copula: Estimated $P_F$ and $P_M$ with $\beta = 0.01$.} \label{table:eq2}
\end{table*}}

\begin{remark}
The asymptotic performance of \eqref{eq:stop11} shows that the average detection time depends on the KL distance provided by each sensor and the KL distance due to the spatial dependence among $L$ sensors. By including the spatial dependence in our analysis, we can reduce the detection time on an average.
\end{remark}

Typically, the copula density function $c_k(\cdot | \bs{\phi}_{k})$  and its corresponding parameter set $\bs{\phi}_{k}$ under hypothesis $H_k, k = 0, 1$ are not known and need to be estimated. Using maximum likelihood estimates in place of the true copula density functions and the true parameters, the copula-based SPRT in \eqref{eq:sprt} becomes a generalized copula-based SPRT.  In the following, we present the estimation of the best copula model. %For high-dimensional data, the defined multivariate copulas, e.g., Gaussian, Student-t, may lack the ability to characterize the potential complex dependence \cite{zhang2019fusion}. In this paper, we adopt a more general copula model,  R-Vine copula, introduced in \cite{bedford2002vines}. 

Since the FC has no knowledge of the dependence structure of the received messages, we assume that the FC waits for $N_0$ messages before starting the copula-based SPRT. Note that $N_0$ can determined by the goodness-of-fit tests for copula models \cite{genest2009goodness}. Hence, the copula density functions and their corresponding parameters can be estimated. The estimation of optimal copula density functions under the two hypotheses is similar. Therefore, the hypothesis index $k$ will be omitted for now to simplify notations. Note that the marginal CDFs need to be evaluated at each time instant.

Before determining the optimal copula, the copula parameter set $\bs{\phi}$ is obtained using MLE, which is given by
\begin{equation}
\label{eq:phi}
\hat{\bs{\phi}} = \arg\max_{ \bs{\phi} } \sum\limits_{i=1}^{N_0}{ \log c(F_{1}(y_{1i}),\ldots, F_{L}(y_{Li})|\bs{\phi}) }.
\end{equation}

Once the copula parameter set is obtained, the best copula $c^*$ is selected from a predefined library of copulas, $\mathcal{C} =  \{c_m: m=1,\ldots,M\}$ using the Akaike Information Criterion (AIC) \cite{akaike1973information} as the criterion, which is given as
\begin{equation}
%c^* = \arg \max_{c_m \in \mathcal{C}} \sum_{i=1}^{N_0}{ \log{c_m( \hat{F}_{1}(v_{1i}),\ldots,\hat{F}_{L}(v_{Li}) | \hat{\bs{\phi}}_m)}}.
\text{AIC}_m =  - \sum_{i = 1}^{N_0} \log c_m(F_{1}(y_{1i}),\ldots, F_{L}(y_{Li}) | \hat{\bs{\phi}}_m) + 2q(m),
\end{equation}
where $q(m)$ is the number of parameters in the $m$th copula model.

The best copula $c^*$ is
\begin{equation}
\label{eq:bestc}
c^* = \arg \min_{c_m \in \mathcal{C}} \text{AIC}_m.
\end{equation}

Now, we have the optimal copula density functions $c_{1}(\cdot | \hat{\bs{\phi}}_{1}^{*})$ and $c_{0}^{*}(\cdot | \hat{\bs{\phi}}_{0}^{*})$ under alternative and null hypotheses, respectively. The log-likelihood ratio test statistics $\Lambda^{n}(\mathbf{y})$ in \eqref{eq:lambda2} is
\begin{equation*}
\begin{aligned}
\Lambda^{n}(\mathbf{y})  = \sum_{i=1}^{n} { \sum_{l=1}^L{ \log \frac{f_{1, l}(y_{li})}{f_{0,l}(y_{li})} }} + \sum_{i=1}^{n}{ \log \frac{c^{*}_1(\mathbf F^{1}(\mathbf y_i) | {\hat{\bs{\phi}}^{*}}_1)}{c^{*}_0( \mathbf F^{0}(\mathbf y_i) | {\hat{\bs{\phi}}^{*}}_0)} }.
\end{aligned}
\label{eq:lambda11}
\end{equation*}

\section{Numerical Results}
\label{sec:nr}
In this section, we demonstrate the efficacy of our proposed copula-based SPRT through numerical examples. There are two hypotheses, where $H_1$ denotes the presence of a signal $s$ and $H_0$ indicates the absence of $s$. Also, $s$ is assumed to be a deterministic signal. We model the signals received at the sensors as:  
\begin{align}
&H_1: z_{li} = s + v_{li}, &l = 1, \ldots L; i = 1, 2, \ldots \\ \nonumber
&H_0: z_{li} = v_{li}, &l = 1, \ldots, L; i = 1, 2, \ldots \nonumber
\end{align}
where $v_{li} \sim \mathcal{N}(0, {\sigma_v^{(l)}}^2)$ is the measurement noise at sensor $l$ and time instant $i$. The received signal $z_{li}$ is assumed to be temporally independent conditioned on either hypothesis. 

The FC receives $y_{li} = z_{li} + w_{li}$, where $w_{li} \sim \mathcal{N}(0, {\sigma_w^{(l)}}^2)$ is the channel noise from the local sensors to the FC. The channel noise is assumed to be temporally independent conditioned on either hypothesis. Also, the channel noise and the signals at the local sensors are mutually independent. However, the received messages $y_{1i}, y_{2i}, \ldots, y_{Li}$ are spatially dependent.

%Therefore, the overall signal-to-noise ratio (SNR) is $6$ dB. 
Unless specified, we assume that $L=3$, $ \sigma_w^l = \sqrt{3}$ and $\sigma_v^l = 1$. We use $N_0 = 100$ observations to estimate the copulas.  The probability of false alarm and miss detection constraints are $\alpha = 0.01$ and $\beta = 0.01$, respectively. The signal spatial dependence is generated using multivariate Gaussian copula. Without loss of generality, we assume that the sensor observations under $H_0$ are independent.
To exhibit the performance improvement by applying our proposed copula-based SPRT, we also evaluate the performance of product-based SPRT that ignores dependence of sensor observations. 

\begin{figure*}[t!]
 \centering
 \begin{minipage}[b]{0.48\linewidth}
  \centering
{\includegraphics[height= 2.3 in, width=3 in]{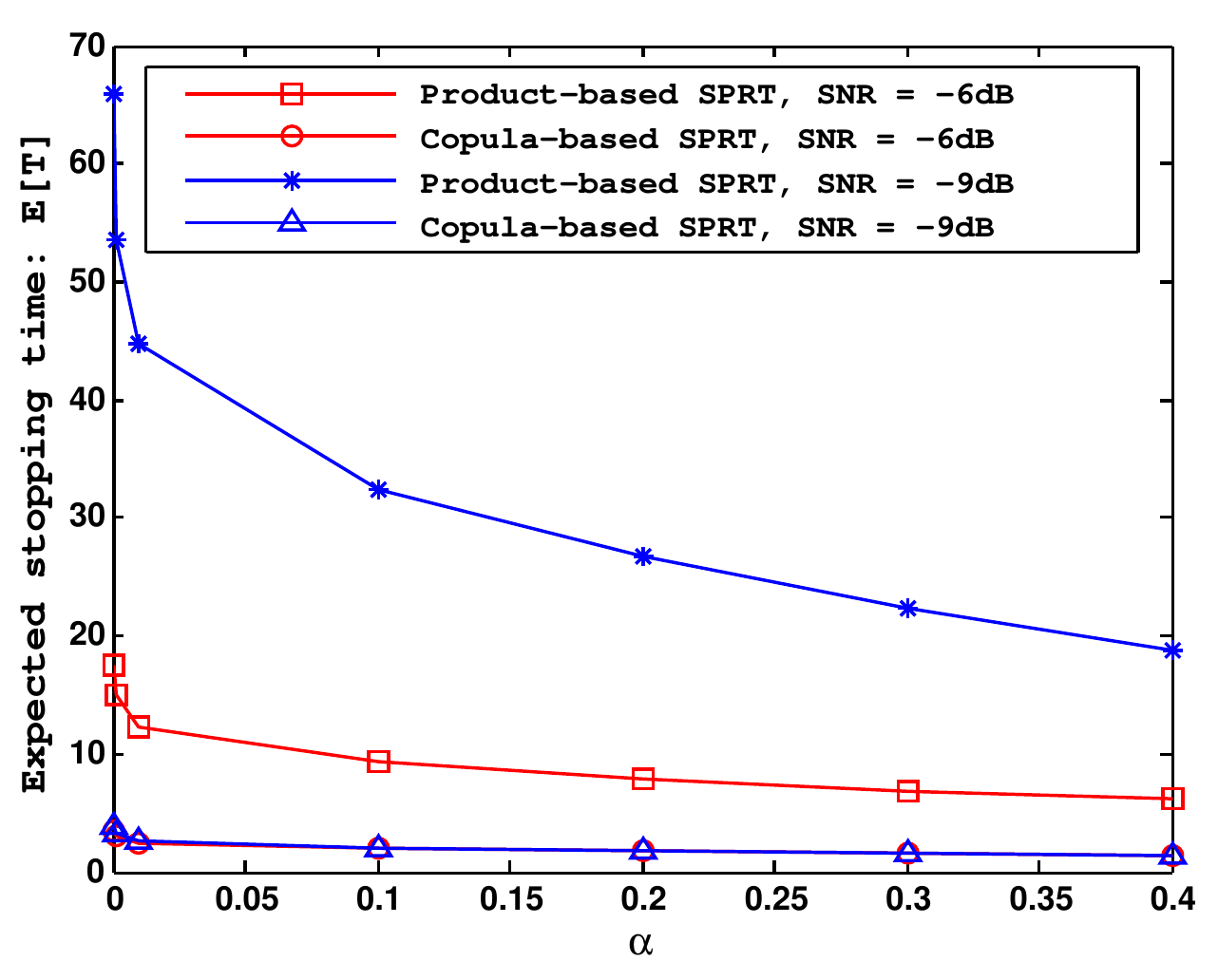} %0.46
\caption{\footnotesize{Expected stopping time as a function of $\alpha$.}} \label{fig:alpha}}
 \end{minipage} 
 \quad
 \begin{minipage}[b]{0.48\linewidth}
  \centering
{ \includegraphics[height= 2.3 in, width=3.2 in]{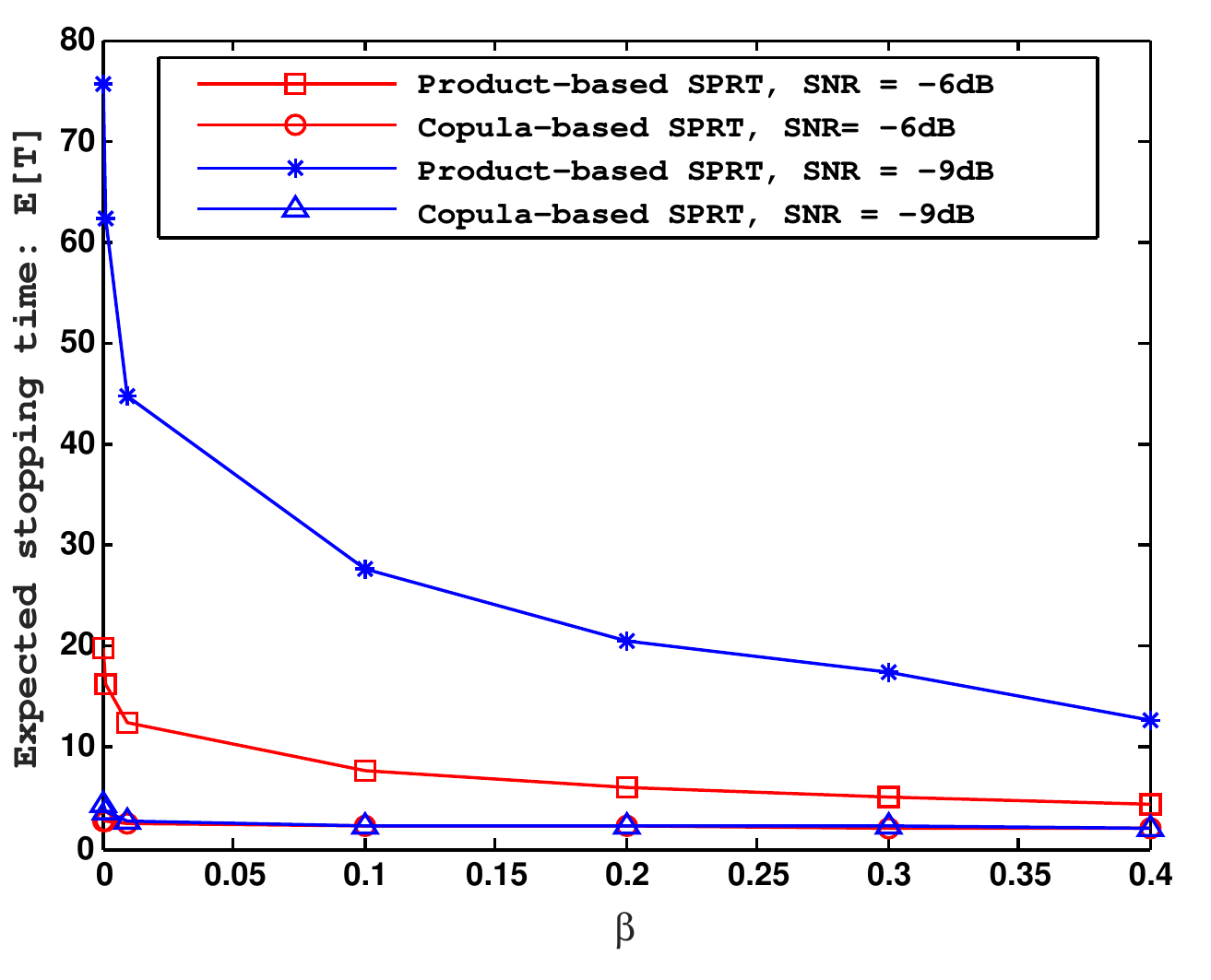} %0.46
 \caption{\footnotesize{Expected stopping time as a function of $\beta$.}}  \label{fig:beta}} 
 \end{minipage}
\end{figure*}

In Table \ref{table:eq1} and Table \ref{table:eq2}, we present the average $P_F$ and $P_M$ values as a function of $\alpha$ and $\beta$, respectively, by comparing the product-based scheme and the copula-based scheme for known copulas and different SNRs. As we can see, the average $P_M$ values obtained for the copula-based SPRT are satisfied given the constraints $\alpha$ and $\beta$ while those for the product-based SPRT are not satisfied.  The average $P_F$ values are satisfied for both the copula-based SPRT and the product-based SPRT. This is because, under $H_0$, we assume that sensor observations are independent. 
Also, in Fig. \ref{fig:alpha} and Fig. \ref{fig:beta}, we show the corresponding expected stopping time $\mathbb E[T]$ with varying $\alpha$ and $\beta$, respectively. As we observe, on average, the copula-based SPRT makes decisions faster than the product-based SPRT. Moreover, for lower SNRs, the product-based SPRT requires more time to complete the detection while the copula-based SPRT is less sensitive to low SNRs.
 
%In Table \ref{table:p-value2}, we present the average $p$ values on the estimation of R-Vine copula models with different $N_0$s.  As we can see, the R-Vine copula model performs very well and it requires at least $N_0 = 30$ observations for $L = 3$ and $N_0 = 40$ for $L= 6, 9$, respectively. 

%	\begin{figure}[t]
%		\centering
%	\includegraphics[height= 2.2 in, width=2.8 in]{alpha_total_2} %0.46
%	\caption{\footnotesize{Expected stopping time as a function of $\alpha$.}}
%	\label{fig:alpha}
%       \end{figure}
%       
%       	\begin{figure}[t]
%		\centering
%	\includegraphics[height= 2.2 in, width=2.8 in]{beta_total_2} %0.46
%	\caption{\footnotesize{Expected stopping time as a function of $\beta$.}}
%	\label{fig:beta}
%       \end{figure}

       	{\renewcommand{\arraystretch}{1.2}
\begin{table}[thb]
\centering
\begin{tabular}{|c|c|c|c|}
\hline
        & $P_F$  & $P_M$ & $\mathbb E[T]$        \\ \hline
Case 1: Product-based SPRT & 0.0061  & 0.1254  & 12.518  \\ \hline
Case 1: Copula-based SPRT  &0.0063  &0.0038 & 2.625   \\ \hline
Case 2: Product-based SPRT & 0.0246  & 0.1254  & 12.317  \\ \hline
Case 2: Copula-based SPRT  &0.0093  &0.0012 & 2.847   \\ \hline
\end{tabular} 
\caption{Unknown-copula:  Estimated $P_F$ and $P_M$ with $\alpha = \beta = 0.01$ and $SNR = -6$ dB.} \label{table:est}
\end{table}}
       
In Table \ref{table:est}, we present the average $P_F$ and $P_M$ values for unknown copula models for two cases. The first case is that only the sensor observations under $H_1$ are dependent, while the second case is that sensor observations under both $H_1$ and $H_0$ are dependent. As we can see,  for the second case, the average $P_F$ and $P_M$ values for the copula-based SPRT are satisfied given the constraints $\alpha$ and $\beta$, respectively, while those for the product-based SPRT are not satisfied. For the first case, the average $P_F$ values are satisfied for the product-based SPRT since under $H_0$, sensor observations are independent. %Moreover, for the second case, the copula-based SPRT needs more time to perform the detection compared to the first case.

%\begin{table}[thb]
%\centering
%\caption{Average $p$ values on the estimation of R-Vine copula model with different number of sensors and $SNR = -6$ dB.}
%\label{table:p-value2}
%\begin{tabular}{|c|c|c|c|}
%\hline
%$N_0$  &$L = 3$ &$L = 6$  & $L = 9$    \\ \hline
%10           &-         & -     & -    \\ \hline
%20          &-      & -     & -         \\ \hline
%30          &  0.911  & -  & -     \\ \hline
%40          &0.938   & 0.999  & 1.000   \\ \hline
%50          &0.953   & 0.998   & 1.000    \\ \hline
%60          &0.926  & 0.999 & 1.000    \\ \hline
%70          &0.939   & 0.999 & 1.000    \\ \hline
%80         &0.939   & 0.998  & 1.000 \\ \hline
%90         &0.945   & 0.999  & 1.000\\ \hline
%100       &0.941  & 0.998  & 1.000 \\ \hline
%120      &0.937  & 0.996   & 1.000\\ \hline
%150     &0.938  & 0.994 & 1.000  \\ \hline
%200     &0.934  & 0.998 & 1.000 \\ \hline
%\end{tabular}
%\end{table}

\section{Conclusion}
\label{sec:conclusion}
In this paper, we proposed a copula-based sequential scheme for the problem of distributed hypothesis testing, where the sensor observations are assumed to be spatially dependent. Moreover, the imperfect communication from the local sensors to the FC was addressed. We have shown the asymptotic optimality of the proposed copula-based SPRT. 
Via simulations, we have shown that our proposed copula-based SPRT can efficiently capture the unknown dependence, and outperform the product-based SPRT which ignores the underlying dependence. Moreover, we have shown that the copula-based SPRT is less sensitive to low SNRs.

\appendices

\bibliographystyle{IEEEtran}
\bibliography{refcopulaJ,ref_seq}
\end{document}